\documentclass[amsmath, amssymb, preprintnumbers, showpacs, showkeys, aps,prx,superscriptaddress,twocolumn]{revtex4-2}
\usepackage{graphicx}
\usepackage{braket}
\usepackage{ulem} 
\usepackage{dsfont}
\usepackage{amsthm,amsmath,amsfonts,amssymb,verbatim,color}
\usepackage{mathtools} 
\usepackage{bbold}
\usepackage{graphicx}
\usepackage[T1]{fontenc}
\usepackage[colorlinks=true,citecolor=blue,linkcolor=blue,urlcolor=blue]{hyperref}
\usepackage{multirow}
\begin{document}
\preprint{}

\title{Complex magnetic phases enriched by charge density waves in topological semimetals GdSb$_x$Te$_{2-x-\delta}$}


\author{Shiming Lei}
\affiliation{Department of Chemistry, Princeton University, Princeton, New Jersey 08544, USA}

\author{Audrey Saltzman}
\affiliation{Department of Physics, Massachusetts Institute of Technology, Cambridge, MA 02139, USA}


\author{Leslie M. Schoop}
\affiliation{Department of Chemistry, Princeton University, Princeton, New Jersey 08544, USA}
\email{lschoop@princeton.edu}

\begin{abstract}
The interplay of crystal symmetry, magnetism, band topology and electronic correlation can be the origin of quantum phase transitions in condensed matter. Particularly, square-lattice materials have been serving as a versatile platform to study the rich phenomena resulting from that interplay. In this work, we report a detailed magnetic study on the square-lattice based magnetic topological semimetal \mbox{GdSb$_x$Te$_{2-x-\delta}$}. We report the \textit{H-T} magnetic phase diagrams along three crystallographic orientations and show that, for those materials where a charge density wave distortion is known to exist, many different magnetic phases are identified. In addition, the data provides a clue to the existence of an antiferromagnetic skyrmion phase, which has been theoretically predicted but not experimentally confirmed in a bulk material yet. 
\end{abstract}

 \date{\today}  
 \maketitle

\section{\label{sec:intro}Introduction}

Square-net materials that exhibit a delocalized, hypervalent chemical bond have been established to be a versatile material platform to host topological electronic states \cite{klemenz2020role}. One interesting class of square-net compounds is the \textit{MXZ} family that crystallizes in the \mbox{PbFCl}-structure, with space group \textit{P4/nmm}. An example is the topological nodal line semimetal \mbox{ZrSiS}. Its crystal structure highlights a square net of Si-atoms, which are separated by a puckered ZiS bilayer \cite{schoop2016dirac}. The band structure near the Fermi level ($E_F$) features linear band-crossings which mostly result from the Si $p_x$- and $p_y$-bands. Without spin-orbit-coupling (SOC), these linear band crossing points (Dirac nodes) connect to form a diamond-shaped loop.

In addition to such nonmagnetic topological materials, magnetic analogs can be also found in the \textit{MXZ} family \cite{Klemenz2019topological}. Particularly, \textit{Ln}SbTe materials (\textit{Ln} = lanthanide) that are isostructural and isoelectronic to ZrSiS have been explored as promising candidates \cite{schoop2018tunable, topp2019effect, hosen2018discovery, sankar2019crystal, weiland2019band, Lei2019Charge, Pandey2020Electronic, yue2020topological}. The introduction of magnetic order can significantly enrich the physics of \textit{MXZ} compounds. For example, it provides an opportunity to access new topological electronic states by breaking time-reversal symmetry; furthermore, the control of magnetic order provides an additional knob to tune the electronic structure and drive it through a manifold of topologically distinct phases \cite{schoop2018tunable}. Such interplay of band topology and magnetism would be especially appealing for spintronic applications, such as topological antiferromagnetic spintronics \cite{vsmejkal2018review}.

Beyond the aforementioned properties, \textit{Ln}SbTe materials also provide a platform to study the physics of charge density waves (CDWs) when the compounds are off-stoichiometric. In the nonmagnetic \mbox{LaSb$_x$Te$_{2-x}$}, DiMasi \textit{et al}. have reported that the CDW wave vector can be continuously tuned along with the Sb composition, \textit{x}, from 0 to 1, which is in accordance with the picture of continuous change of band filling and Fermi surface nesting \cite{dimasi1996stability}. As an extension, we recently explored the antiferromagnetic \mbox{GdSb$_x$Te$_{2-x-\delta}$} series ($\delta$ indicates the vacancy-level) and found that this system can be tuned to have either commensurate or incommensurate lattice modulations depending on the Sb content. Compared to \mbox{LaSb$_x$Te$_{2-x}$}, where CDW appears when $x<1$, in \mbox{GdSb$_x$Te$_{2-x-\delta}$} the non-distorted tetragonal phase is stable for $0.85<x<1$ \cite{Lei2019Charge}.

Based on the observation of CDWs and band crossings protected by non-symmorphic symmetry \cite{young2015dirac,schoop2016dirac}, we suggested a strategy for using the CDW as a tool to create idealized non-symmorphic Dirac semimetals in the \textit{Ln}SbTe phases \cite{lei2020CDW}. We showed that the electronic structure of \mbox{GdSb$_{0.46}$Te$_{1.48}$} near $E_F$ features clean non-symmorphic nodal-line Dirac states, while other states at $E_F$ are suppressed by CDWs. Since the magnetic susceptibility is affected by the Fermi energy in nodal-line semimetals \cite{koshino2016magnetic,mikitik2016magnetic}, it would be of particular interest to understand the evolution of the magnetism with varying band filling (in this case, varying Sb content) in the \mbox{GdSb$_x$Te$_{2-x-\delta}$} system.



Here, we report a systematic study of the magnetic properties of \mbox{GdSb$_x$Te$_{2-x-\delta}$}, with a focus on the compounds with CDW distortions (x < 0.85). By magnetic susceptibility decoupling and mechanical detwinning, we are able to establish \textit{H-T} magnetic phase diagrams along all three crystallographic orientations. The decoupled magnetic properties show clear in-plane magnetic anisotropy, pointing to the important role of the unidirectional CDW distortion in breaking the $C_4$ crystal symmetry. Based on the magnetic properties of eleven different compositions with \textit{H//c}, we also report an Sb-composition dependent phase diagram. Interestingly, the idealized non-symmorphic Dirac semimetal \mbox{GdSb$_{0.46}$Te$_{1.48}$} is found to exhibit an anomalously enhanced magnetic susceptibility in a narrow \textit{H-T} phase regime. We discuss the possibility that this is a signature of an antiferromagnetic skrymion phase. Overall, the rich magnetic phase diagrams are linked to the magnetocrystalline anisotropy and its delicate competition with other energy contributions. In order to understand the role of the out-of-plane magnetic exchange interaction, we also report the magnetic phase diagram of the antiferromagnetic van der Waals material \mbox{GdTe$_3$}, whose crystal structure features a similar CDW distorted square-net lattice and puckered GdTe bilayers, but exhibits fewer and weaker out-of-plane exchange couplings.  

\section{Methods}
Single crystals of \mbox{GdSb$_x$Te$_{2-x-\delta}$} were grown by chemical vapor transport, using iodine as the transport agent. For more details of the growth conditions, see Ref. \cite{Lei2019Charge}. By varying the amount of Sb in the starting materials, a series of \mbox{GdSb$_x$Te$_{2-x-\delta}$} single crystals with varying Sb content was obtained. The elemental composition was analyzed by energy dispersive x-ray spectroscopy (EDX) using a Verios 460 Scanning Electron Microscope with an Oxford energy dispersive x-ray spectrometer and with incident electron energy of 15 keV. Single crystals of \mbox{GdTe$_3$} were grown by a self-flux method with excess of tellurium. The growth condition was the same as described in Ref. \cite{lei2020high}. The magnetic properties were analyzed by a Quantum Design PPMS DynaCool system via the vibrating sample magnetometer (VSM) option. 


\section{Results and Discussion}

\subsection{\label{subsec1} Magnetic behavior of \mbox{GdSb$_{0.46}$Te$_{1.48}$}}

\begin{figure}
\includegraphics[width=0.4\textwidth]{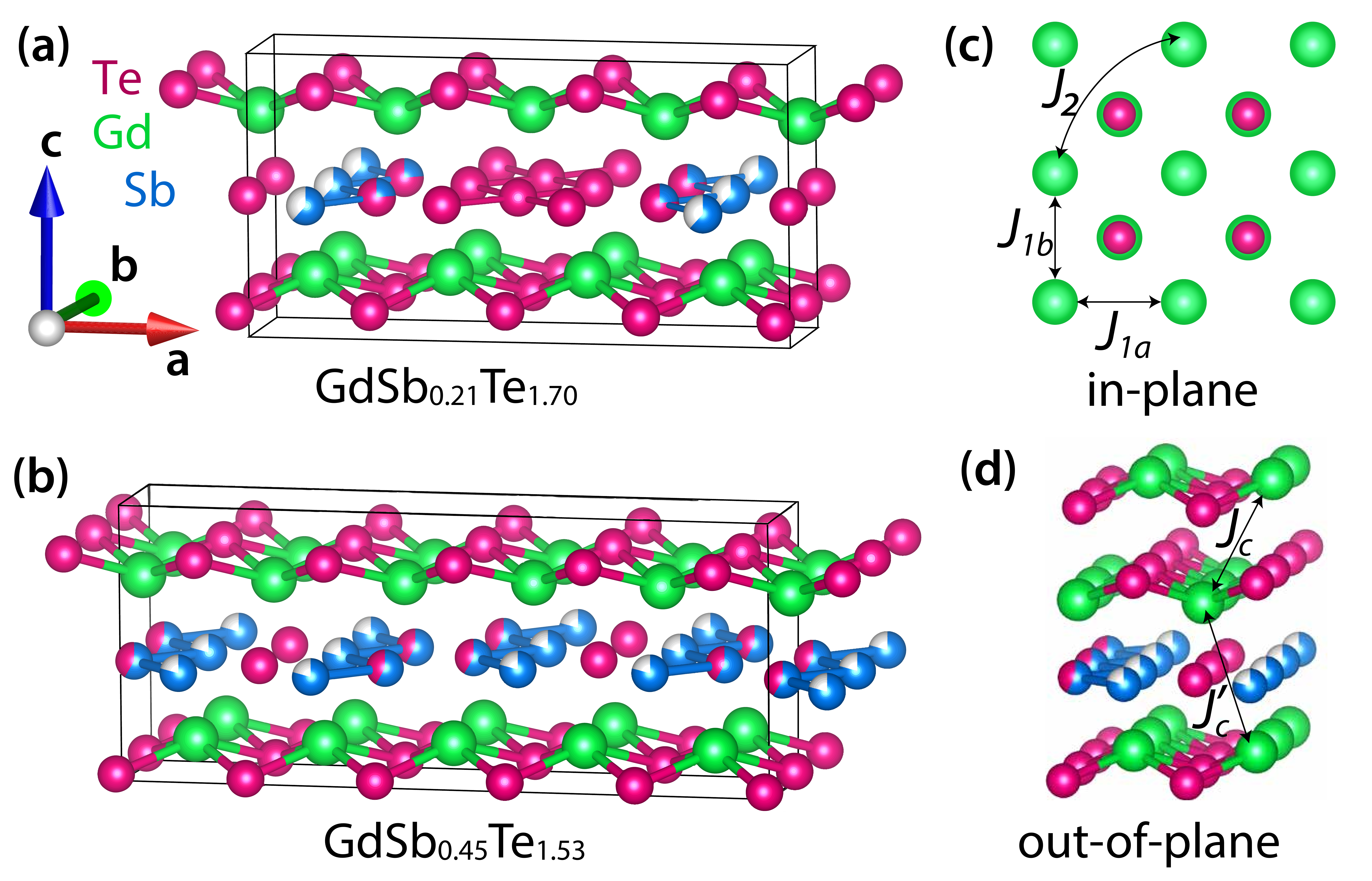}
\caption{\label{fig:structure} Two different structures that can be observed in \mbox{GdSb$_x$Te$_{2-x-\delta}$} depending on composition \cite{Lei2019Charge}, as well as the responsible magnetic exchange interactions in a simplified structure. Crystal structures of \mbox{GdSb$_{0.21}$Te$_{1.70}$} (a) and \mbox{GdSb$_{0.45}$Te$_{1.53}$} (b), with four- and five-fold superstructural modulations, respectively. Solid lines represent the unit cells. The superstructural modulation appears along the \textit{a}-axis. Crystal structure solutions are from Ref. \cite{Lei2019Charge}. In-plane (c) and out-of-plane (d) magnetic exchange couplings as they appear in a simplified orthorhombic structure, when ignoring the more complex supercell modulations. Note that \textit{J$_{1a}$} and \textit{J$_{1b}$} (nearest-neighbor interactions), and \textit{J$_{2}$} (next-nearest-neighbor interaction) are illustrated as in-plane couplings, while \textit{J$_{c}$} (\textit{nn}) and \textit{$J^\prime_{c}$} (\textit{nnn}) as out-of-plane couplings.}
\end{figure}

Two examples of \mbox{GdSb$_x$Te$_{2-x-\delta}$} superstructures resolved by single-crystal x-ray diffraction for $x<0.85$ are shown in Figs. ~\ref{fig:structure}(a) and (b), which correspond to \mbox{GdSb$_{0.21}$Te$_{1.70}$} and \mbox{GdSb$_{0.45}$Te$_{1.53}$}, respectively \cite{Lei2019Charge}. Our prior study indicated that the measured effective magnetic moment for all \mbox{GdSb$_x$Te$_{2-x-\delta}$} is close to the theoretical free-ion value of 7.94 $\mu_B$ for Gd$^{3+}$ with $4f^7$ configuration \cite{Lei2019Charge}. Therefore, the ground state angular momentum of Gd is \textit{J} = 7/2 (\textit{S} = 7/2, \textit{L} = 0).
For the simplification of the discussion, we assume the Gd lattice to be of simple rectangular symmetry. Then, at least two nearest-neighbor (\textit{nn}) magnetic exchange interaction terms, \textit{J$_{1a}$} and \textit{J$_{1b}$}, and one next-nearest-neighbor (\textit{nnn}) term, \textit{J$_{2}$}, are expected to be responsible for the in-plane magnetic coupling. Since the structure also exhibits two puckered GdTe layers, which are stacked along the \textit{c}-axis [Fig. ~\ref{fig:structure}(d)], a \textit{nn} out-of-plane coupling, \textit{J$_{c}$}, and a \textit{nnn} out-of-plane coupling, $J^\prime_{c}$, also play a role, as we will discuss below. As mentioned in the introduction, \mbox{GdSb$_{0.46}$Te$_{1.48}$} was recently found to be a nearly ideal non-symmorphic Dirac semimetal with an almost ``clean'' Dirac node at the Fermi level \cite{lei2020CDW}. For this reason, we begin our discussion with the magnetic behavior of \mbox{GdSb$_{0.46}$Te$_{1.48}$}.

\begin{figure}
\includegraphics[width=0.5\textwidth]{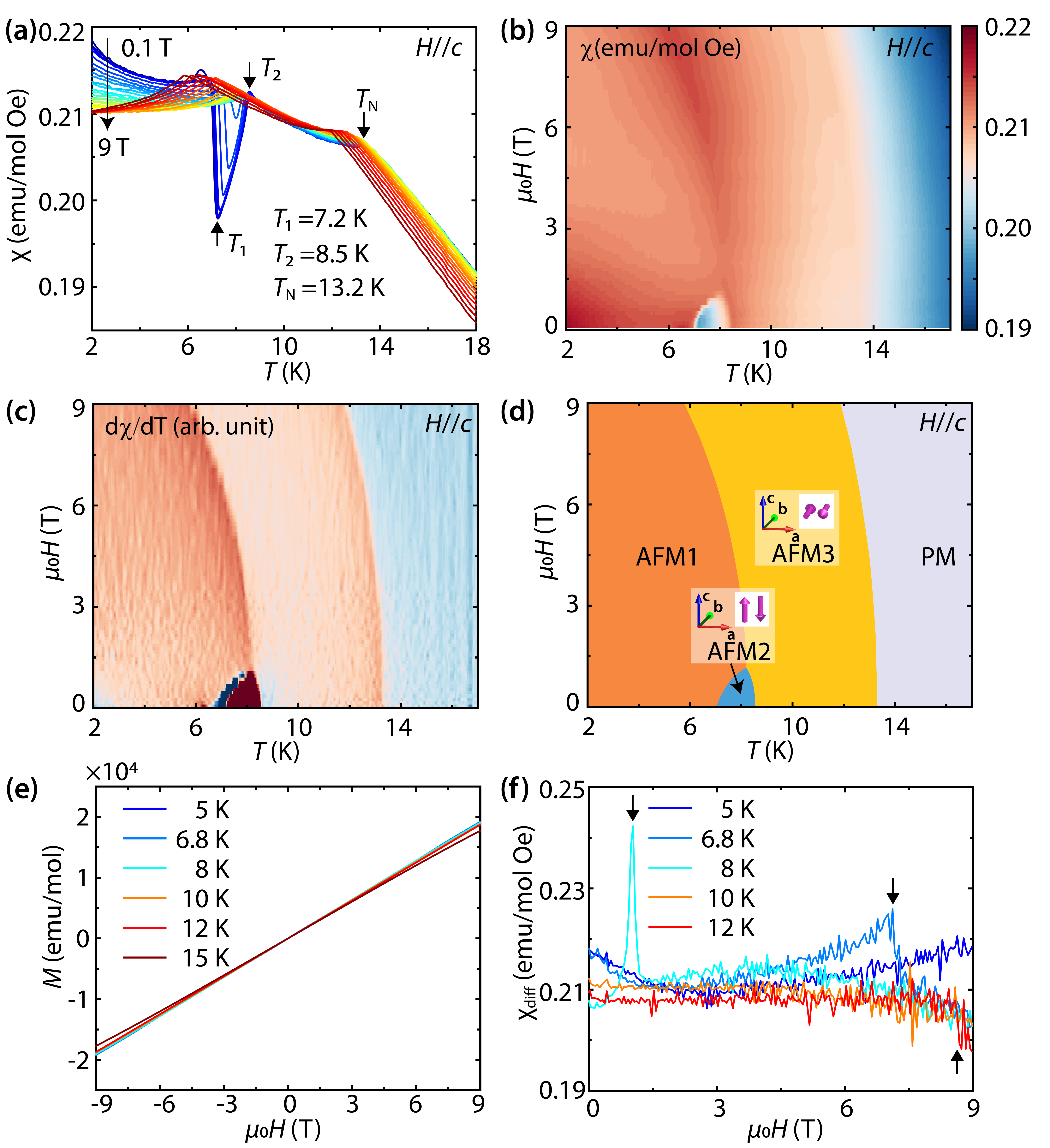}
\caption{\label{fig:GdSb0.46_Hc}Magnetic phase diagram of \mbox{GdSb$_{0.46}$Te$_{1.48}$} for \textit{H//c}. (a) Temperature dependent DC magnetic susceptibility ($\chi$) at selected field range from 0.1\,T to 9\,T. The three magnetic transitions at 0.1\,T are indicated by the arrows. (b) Map of the DC magnetic susceptibility in the parameter space of magnetic field and temperature. (c) The temperature derivative of the magnetic susceptibility (d$\chi$/d$T$) shows clear magnetic phase boundaries. (d) The magnetic phase diagram inferred from magnetic susceptibility measurements. PM indicate the paramagnetic phase. AFM1, AFM2, AFM3 are antiferromagnetic phases. The inferred spin orientations for AFM2 and AFM3 phases are also illustrated. (e) Sampled \textit{MH}-data at selected temperatures. (f) The differential magnetic susceptibility, $\chi_{\text{diff}}$ = d\textit{M}/d\textit{H}, derived from the \textit{MH}-data shown in (e). The field induced magnetic transitions are indicated by the arrows for the data measured at \textit{T} = 6.8\,K, 8\,K, and 12\,K. Note that the magnetic susceptibility data from 2\,K to 16\,K was reported in Ref. \cite{lei2020CDW}, and is replotted here in (b) for the convenience of discussion.}
\end{figure}

Figure ~\ref{fig:GdSb0.46_Hc} shows the magnetic properties of \mbox{GdSb$_{0.46}$Te$_{1.48}$} for \textit{H//c}. The magnetic susceptibility for $\mu_0$\textit{H} = 0.1\,T reveals three magnetic transitions, $T_1 =\,$7.2\,K, $T_2 =\,$8.5\,K, and $T_N =\,$13.2\,K. Under increasing magnetic field, the temperature window between the $T_1$ and $T_2$ transitions gradually shrinks and disappears at a critical field of 1.1\,T. Upon further field increase, the magnetic transitions shift to lower temperatures. Figure ~\ref{fig:GdSb0.46_Hc}(e) shows the \textit{MH}-data for field sweep between -9\,T and 9\,T. The almost linear field-dependent \textit{MH}-curves and zero remanent magnetization suggest that the three magnetic ordered phases, AFM1, AFM2, and AFM3 are antiferromagnetic. The field-induced magnetic phase transitions are better revealed in the differential  magnetic susceptibility curves [Fig.~\ref{fig:GdSb0.46_Hc}(f)].

\begin{figure}
\includegraphics[width=0.5\textwidth]{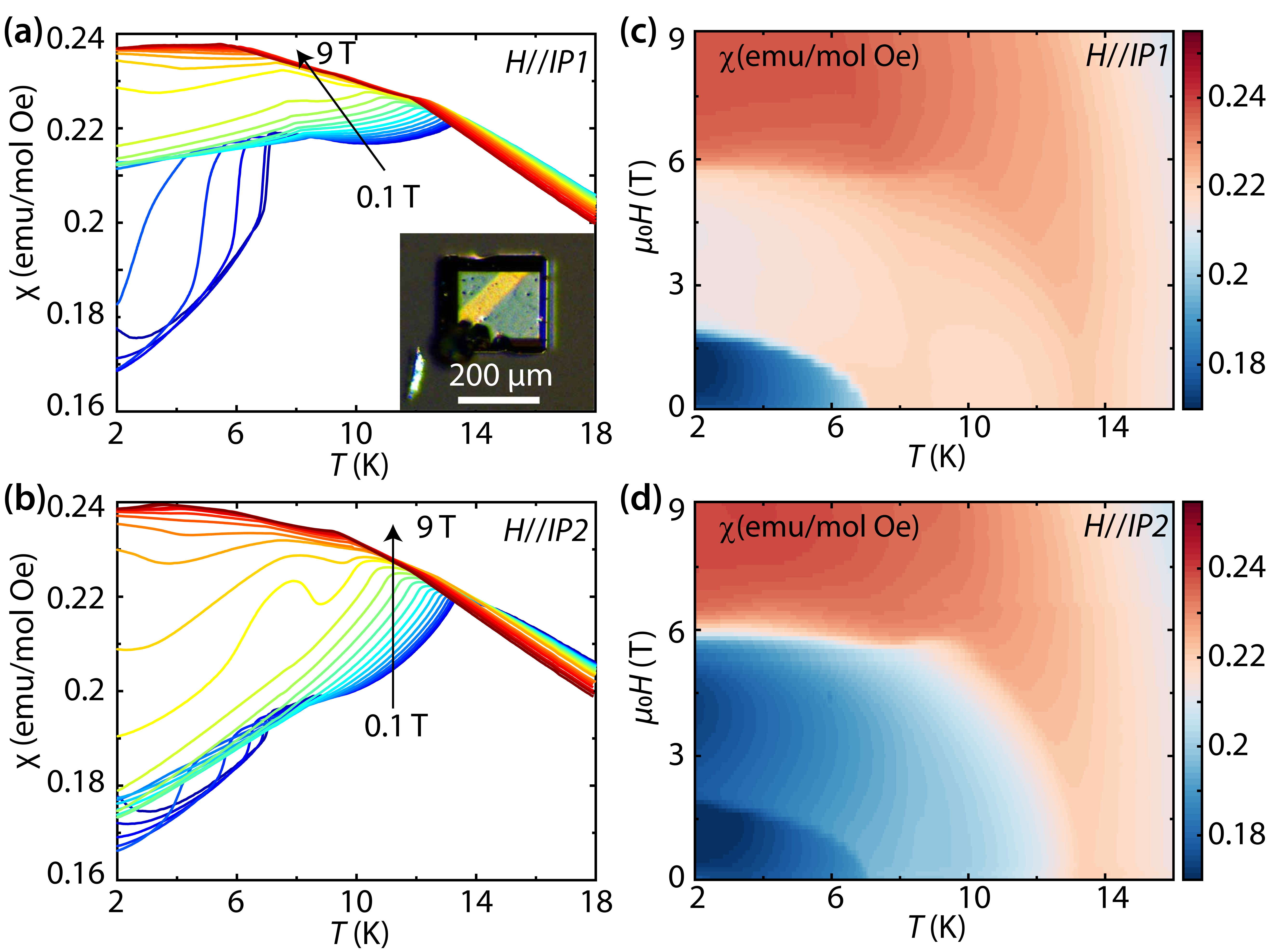}
\caption{\label{fig:GdSb0.46_Ha_Hb}In-plane magnetic anisotropy of \mbox{GdSb$_{0.46}$Te$_{1.48}$}. The temperature dependent DC magnetic susceptibility ($\chi$) at selected field range from 0.1\,T to 9\,T for the two orthogonal in-plane orientations, \textit{IP1} (a) and \textit{IP2} (b). The inset in (a) shows a polarized optical image of a \mbox{GdSb$_{0.46}$Te$_{1.48}$} crystal. The bright-dark contrast indicates the two types of CDW domains, whose CDW wave vectors are orthogonal to each other. (c) and (d) DC magnetic susceptibility maps corresponding to the two in-plane orientations.}
\end{figure}

To better understand these magnetic phases, we also explored the magnetic properties for fields parallel to the \textit{ab}-plane. Since a magnetocrystalline anisotropy is expected for the two in-plane crystallographic axes (\textit{a}- and \textit{b}-axis shown in Fig. \ref{fig:structure}), due to the unidirectional CDW distortion, we performed two sets of independent magnetization measurements for either \textit{H//IP1} or \textit{H//IP2}. The results are shown in Figs. \ref{fig:GdSb0.46_Ha_Hb}(a) and (b). \textit{IP1} and \textit{IP2} are aligned to be either parallel or perpendicular to \textit{a}- and \textit{b}-axes, respectively. For crystals that undergo a tetragonal to orthorhombic structural transition from high to low temperature, ferroelastic structural twinning is often observed in the lower-symmetry phase.  Examples for this are the iron arsenide superconductors \mbox{\textit{A}Fe$_2$As$_2$} (\textit{A}\,=\,Ca, Sr and Ba) \cite{tanatar2010uniaxial,chu2010plane}, and \mbox{YBa$_2$Cu$_3$O$_{7-x}$} single crystals \cite{giapintzakis1989method}. The same happens in the orthorhombic \mbox{GdSb$_x$Te$_{2-x-\delta}$} system. The inset of Fig. \ref{fig:GdSb0.46_Ha_Hb}(a) shows a polarized optical image of a \mbox{GdSb$_{0.46}$Te$_{1.48}$} crystal with clear ferroelastic twinning. Magnetization measurements on such crystals with fields aligned along either \textit{H//IP1} or \textit{H//IP2} consist of intrinsic magnetic responses from twin domains in which both \textit{H//a} and \textit{H//b} contribute; the weight of each intrinsic response depends on their relative domain volume ratio. If the volume of one type of domain is predominant over the other, the in-plane magnetic anisotropy can still be resolved. The temperature dependent magnetic susceptibility of \mbox{GdSb$_{0.46}$Te$_{1.48}$} for \textit{H//IP1} [Fig.~\ref{fig:GdSb0.46_Ha_Hb}(a)] and \textit{H//IP2} [Fig.~\ref{fig:GdSb0.46_Ha_Hb}(b)] shows a clear difference below \textit{T}$_N$\,=\,13.2\,K at low field. The difference is also visualized in the magnetic susceptibility map [Figs.~\ref{fig:GdSb0.46_Ha_Hb}(c) and (d)].


\begin{figure*}
\includegraphics[width=1\textwidth]{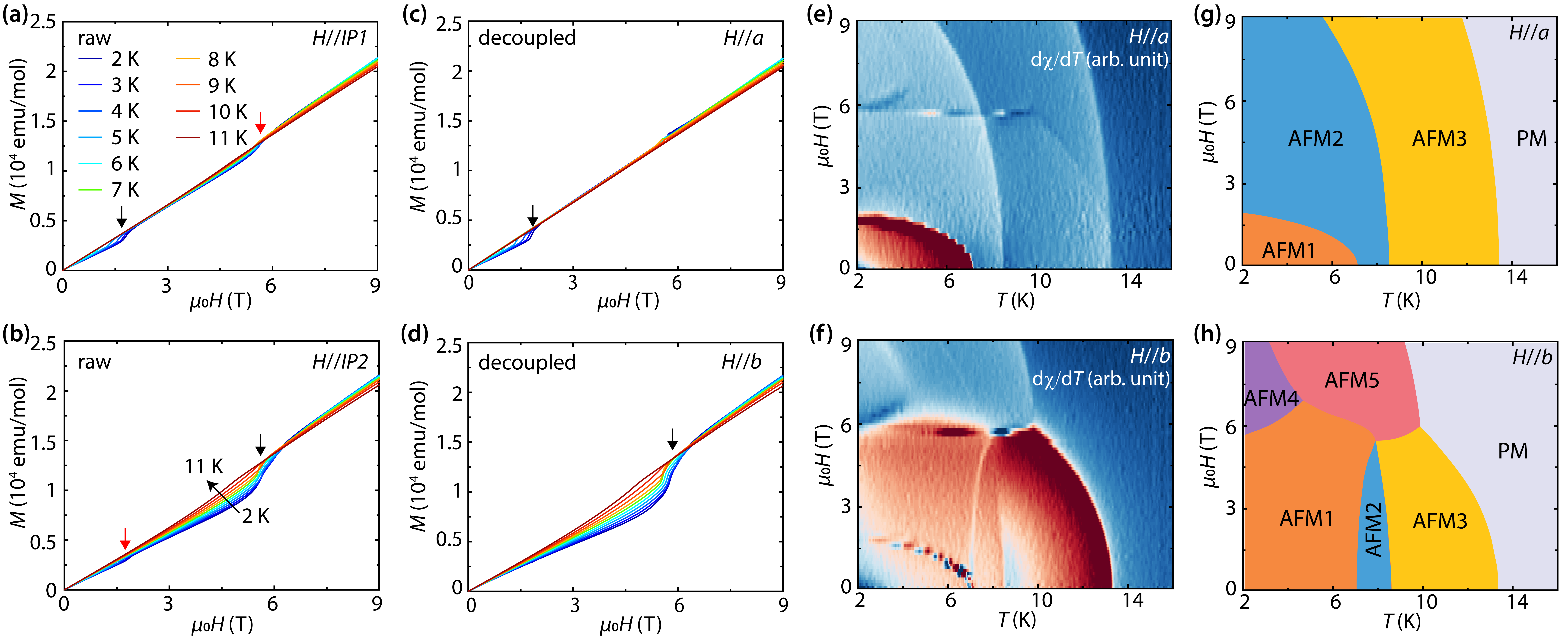}
\caption{\label{fig:GdSb0.46_phase_Ha_Hb}Magnetic phase diagram of \mbox{GdSb$_{0.46}$Te$_{1.48}$} for fields parallel to the two in-plane orientations. \textit{MH}-measurements for \textit{H//IP1} (a) and \textit{H//IP2} (b), respectively, from 2\,K to 11\,K. The field induced magnetic transitions are indicated by arrows. The red arrows indicate the weaker transition feature resulting from domain mixing. Decoupled \textit{MH}-data for \textit{H//a} (c) and \textit{H//b} (d), respectively. (e) and (f) Temperature derivatives of the decoupled magnetic susceptibilities. (g) and (h) The magnetic phase diagrams for \textit{H//a} and \textit{H//b}, respectively. PM indicates the paramagnetic phase. AFM1, AFM2, AFM3, AFM4, and AFM5 are antiferromagnetic phases, as discussed in the main text.}
\end{figure*}

To gain a better understanding on the magnetic anisotropy, we performed further \textit{MH}-measurements for fields parallel to \textit{IP1} and \textit{IP2} [Figs. \ref{fig:GdSb0.46_phase_Ha_Hb}(a) and (b)]. Two spin-flop transitions can be observed at $\mu_0H\approx  1.7\,$T and 5.7\,T in the curves measured at $T$ = 2\,K and 3\,K, as indicated by the arrows. For measurements on a single crystal that would only consist of a single domain, the intrinsic \textit{MH}-data can be formally described as a function, $M_a$(\textit{H}) for \textit{H//a} and $M_b$(\textit{H}) for \textit{H//b}. If mixed domains are present, the measured magnetization can be considered as a simple summation of the magnetization for the two species of domains (the negligible response from domain wall region is ignored),  
\begin{equation} \label{eq:1}
M_{IP1}\left(H\right) =  \lambda M_a\left(H\right) +  \left(1-\lambda\right)M_b\left(H\right)
\end{equation}
\begin{equation} \label{eq:2}
M_{IP2}\left(H\right) =  \lambda M_b\left(H\right) +  \left(1-\lambda\right)M_a\left(H\right)
\end{equation}
where $M_{IP1}$ and $M_{IP2}$ denote the magnetization measured along \textit{IP1} and \textit{IP2} orientations, respectively, and $\lambda$ describes the volume fraction of the \textit{a}-domain, in which the \textit{a}-axis is parallel to \textit{IP1} orientation. For  $\lambda=1$, the measurement for \textit{H//IP1} is equivalent to \textit{H//a}. From eqs. (1) and (2), the intrinsic magnetic response $M_a$(\textit{H}) and $M_b$(\textit{H}) can be derived:
\begin{equation} \label{eq:3}
M_a\left(H\right) =  \frac{-\lambda}{1-2\lambda}M_{IP1}\left(H\right) + \frac{1-\lambda}{1-2\lambda}M_{IP2}\left(H\right)
\end{equation}
\begin{equation} \label{eq:4}
M_b\left(H\right) = \frac{1-\lambda}{1-2\lambda}M_{IP1}\left(H\right) + \frac{-\lambda}{1-2\lambda}M_{IP2}\left(H\right)
\end{equation}

Once $\lambda$ is known, the intrinsic magnetic response without a domain mixing effect can be obtained even though the measurements were performed on a crystal with CDW twin domains. Here we utilize the features of the field-induced spin-flop transitions at $\mu_0H\approx\,1.7\,$T and 5.7\,T [Fig. \ref{fig:GdSb0.46_phase_Ha_Hb}(a) and (b)] to evaluate $\lambda$. A detailed description of the \textit{MH}-data decoupling is provided in the Supplemental Material. The intrinsic \textit{MH}-response after decoupling is shown in Figs. \ref{fig:GdSb0.46_phase_Ha_Hb}(c) and (d).

Up to this stage, we have obtained two sets of intrinsic \textit{MH}-data corresponding to field conditions of \textit{H//a} and \textit{H//b}. However, which set of \textit{MH}-curves corresponds to which field orientation is still unknown. To address this, we took advantage of mechanical detwinning. For orthorhombic crystals with in-plane ferroelastic twinning, external mechanical stress can tune the domain population \cite{zheng2018bulk, Lei2018Observation}. For orthorhombic \mbox{GdSb$_x$Te$_{2-x-\delta}$}, the subcell lattice parameter, $a$, along the CDW wave vector direction, is larger than the other in-plane parameter, \textit{b}. Therefore, uniaxial compressive stress along either in-plane crystallographic axis is able to facilitate the preferential growth of domains that have the short $b$-axis along the stress direction. Such mechanical approach has been used before to detwin \mbox{\textit{A}Fe$_2$As$_2$} (\textit{A}\,=\,Ca, Sr and Ba) single crystals \cite{tanatar2010uniaxial,chu2010plane}. With mechanical detwinned crystals, we can conclude that the spin-flop transitions at $\mu_0H\approx  1.7\,$T and 5.7\,T correspond to \textit{H//a} and \textit{H//b}, respectively, as indicated in Fig. \ref{fig:GdSb0.46_phase_Ha_Hb}(c) and (d). More details on the mechanical twinning are provided in Supplemental Material. Using the same \textit{MH}-decoupling procedure as described above, we could derive the resulting magnetic phase diagrams for \textit{H//a} and \textit{H//b} [(Figs. \ref{fig:GdSb0.46_phase_Ha_Hb}(e)-(f)]. 

We now discuss the orientation of the magnetic moment in the different magnetic ordered phases. Since Gd strongly absorbs neutrons, it is difficult to determine the magnetic structure with neutron diffraction. Nevertheless, we are able to gain some insight on the nature of the magnetic phases based on the magnetic susceptibility measured along the three principal crystallographic orientations [$\chi_c$ in Fig. \ref{fig:GdSb0.46_Hc}(a), decoupled $\chi_a$ and  $\chi_b$ in Fig.  S2 in the Supplemental Material]. At \textit{T}$_N$, there is a peak in $\chi_b$, while $\chi_c$ and $\chi_a$ become relatively flat for $T_2$\,<\,\textit{T}\,<\,$T_N$, thus the spins in the AFM3 phase are clearly oriented along \textit{b}-axis [illustrated in Fig. \ref{fig:GdSb0.46_Hc}(d)]. 
For $T_1$\,<\,\textit{T}\,<\,$T_2$, there is a sharp drop in $\chi_c$, in contrast to a subtle change in the slope of $\chi_a$ and $\chi_b$. This implies a spin-reorientation transition to a new antiferromagnetic phase [AFM2, illustrated in Fig. \ref{fig:GdSb0.46_Hc}(d)], with the spins flopping from the \textit{b}- to the \textit{c}-axis. Below \textit{T}$_1$, $\chi_c$ sharply recovers from the dip and remains relatively flat down to the lowest measured temperature, while $\chi_a$ sharply drops, and both $\chi_a$ and $\chi_b$ continuously decrease upon cooling. This suggests another spin-reorientation transition into an antiferroamgnetic state with the spins flopping from \textit{c}- to the \textit{ab}-plane. Since $\chi_a$ and $\chi_b$ show similar temperature dependent behavior down to the lowest measured temperature, the magnetic easy axis of the AFM1 phase may point to an in-plane orientation other than \textit{a}- or \textit{b}-axes, or the AFM1 phase exhibits a spiral spin texture with magnetic moments lying in the \textit{ab}-plane.

At higher fields, field-induced magnetic phases, AFM4 and AFM5, appear for \textit{H//b} [Fig. \ref{fig:GdSb0.46_phase_Ha_Hb}(h)]. Since the \textit{MH}-curves are still linear and the extrapolation of the magnetization down to zero field is nearly zero, we conclude that AFM4 and AFM5 are also antiferromagnetic phases. Considering the characteristic spin-flop transitions at $\mu_0H\approx5.7\,$T, both AFM4 and AFM5 phases are likely to have spins oriented along \textit{a}-axis, although their respective antiferromagnetic spin configurations are different. 

\begin{figure}
\includegraphics[width=0.5\textwidth]{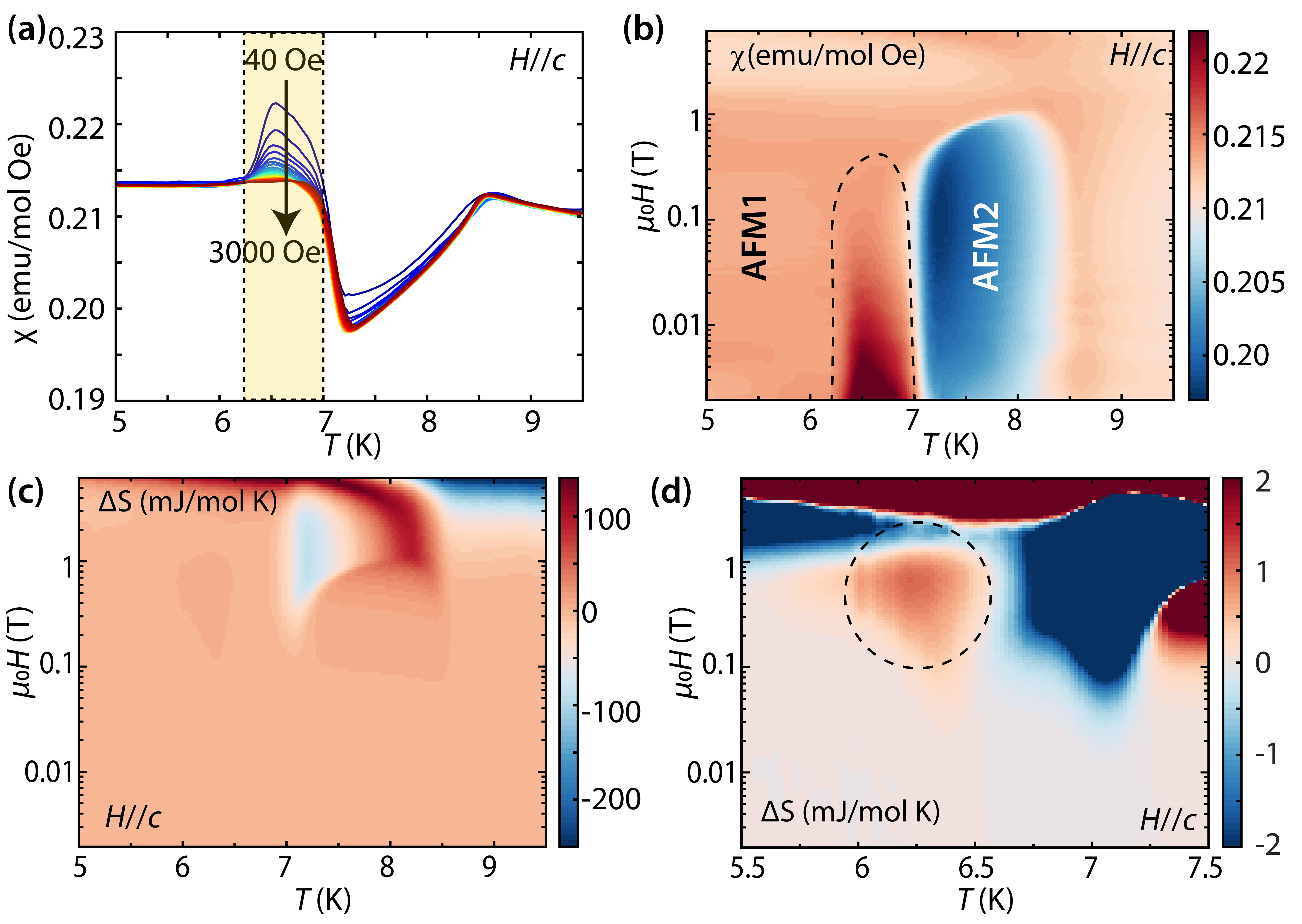}
\caption{\label{fig:GdSb0.46_low_field}Low-field magnetic behavior of \mbox{GdSb$_{0.46}$Te$_{1.48}$} for \textit{H//c}.  (a) Temperature dependent DC magnetic susceptibility ($\chi$) at a selected field range from 40\,Oe to 3000\,Oe. The temperature window where $\chi$ shows anomalous enhancement is indicated by the shaded rectangle. (b) The low-field magnetic susceptibility map. The region with enhanced magnetic susceptibility is outlined by the dashed line. (c) Map of the relative magnetic entropy, $\Delta$\textit{S}. (d) Zoom-in re-scaled relative magnetic entropy. The region with enhanced $\Delta$\textit{S} is indicated by the dashed circle. Note that the vertical axes in (b-d) are plotted in a log scale.}
\end{figure}

We next focus on the low-field behavior of \mbox{GdSb$_{0.46}$Te$_{1.48}$}. Figure \ref{fig:GdSb0.46_low_field}(a) shows $\chi_c(T)$ for a field range from 40\,Oe to 3000\,Oe. An anomalously enhanced $\chi_c$ is observed in the temperature range of 6.2\,K <\,\textit{T}\,<\,7.0\,K. The anomalous behavior is better revealed in the magnetic susceptibility map [Fig. \ref{fig:GdSb0.46_low_field}(b)], where the anomalous region is surrounded by the AFM1 and AFM2 phases. Since AFM1 and AFM2 phases were concluded to have an in-plane and out-of-plane spin orientation, respectively, the phase regime in between is of particular interest: the in-plane and out-of-plane magnetic anisotropy energy are expected to have a delicate balance. A non-collinear spin texture could be a likely consequence. One possibility could be the appearance of skyrmions \cite{muhlbauer2009skyrmion, yu2011near, onose2012observation}, which are particle-like spin textures of topological origin.

One signature of a skyrmion phase can be revealed by a magnetoentropic analysis, as demonstrated by the recent studies on \mbox{FeGe} \cite{bocarsly2018magnetoentropic} and \mbox{Co$_8$Zn$_9$Mn$_3$} \cite{bocarsly2019deciphering}. Formally, the magnetic 
entropy change $\Delta$$S_M$$\left(T, H\right)$ is expressed as:
\begin{equation} \label{eq:5}
\begin{aligned}
\Delta S_M\left(T, H\right) & = S_M\left(T, H\right) - S_M\left(T, 0 \right) \\
&\quad= \int_{0}^{H} \left[\frac{\partial S\left(T, H\right)}{\partial H}\right]_T \text{d}H,
\end{aligned}
\end{equation}
where $S$ is the total entropy. With Maxwell’s relation \(\left[\partial S(T, H)/\partial H\right]_T = \left[\partial M(T, H)/\partial T\right]_H\), the equation above can be rewriten as:
\begin{equation} \label{eq:6}
\begin{aligned}
\Delta S_M\left(T, H\right) = \int_{0}^{H} \left[\frac{\partial M(T, H)}{\partial T}\right]_H \text{d}H.
\end{aligned}
\end{equation}

Therefore, the magnetocaloric response of a skyrmion host can be experimentally evaluated by a series of $M\left(T\right)$ measurements under different applied magnetic fields. To visualize the field-induced isothermal magnetic entropy change in the \textit{H-T} space, the entropy \(\Delta S_M\left(T, H \right)\) at zero field is considered as a baseline and \(\Delta S_M\left(T, 0 \right)\) is set to zero. Figure \ref{fig:GdSb0.46_low_field}(c) shows the map of \(\Delta S_M\left(T, H \right)\) obtained using eq. (6). Figure \ref{fig:GdSb0.46_low_field}(d) shows the re-scaled zoom-in view, highlighting the enhanced entropy region (outlined by the dashed circle). Such a feature is similar to that observed in \mbox{FeGe} \cite{bocarsly2018magnetoentropic} and \mbox{Co$_8$Zn$_9$Mn$_3$} \cite{bocarsly2019deciphering}, with an enhanced entropy in the skyrmion lattice regime relative to the neighboring phases.

The existence of skyrmions in magnetic materials is usually a result of a delicate interplay between different energy terms \cite{zhang2020skyrmion}. So far the majority of known skyrmion-hosting materials are noncentrosymmetric, and the critical energy term that is responsible for the non-collinear spin texture of skyrmions is the Dzyaloshinskii-Moriya (DM) interaction \cite{dzyaloshinsky1958thermodynamic, moriya1960anisotropic}.


However, it has been theoretically proposed that frustrated exchange interactions \cite{okubo2012multiple, leonov2015multiply, lin2016ginzburg} and higher-order coupling beyond the Ruderman-Kittel-Kasuya-Yosida (RKKY) interaction \cite{ozawa2017zero, hayami2017effective} are two possible mechanisms to introduce skyrmions in a centrosymmetric material. For the former mechanism, both triangular and square lattices were discussed due to the competition of \textit{nn}-interaction and further-\textit{nn}-interactions. For the latter, the effective multiple spin coupling beyond the conventional RKKY interaction is the key ingredient for stabilization of the skyrmion phase in itinerant electron systems with local magnetic moments. Soon after these theoretical proposals, the emergence of a skyrmion state in a centrosymmetric material was experimentally discovered in the frustrated metallic magnets, \mbox{Gd$_2$PdSi$_3$} (Gd forms a layered triangular lattice)  \cite{kurumaji2019skyrmion},  \mbox{Gd$_3$Ru$_4$Al$_{12}$} (Gd forms a layered distorted kagom\'e lattice) \cite{hirschberger2019skyrmion}, and \mbox{GdRu$_2$Si$_2$} (Gd forms a layered square net) \cite{khanh2020nanometric}. The origin of skyrmions in these materials was linked to these proposed mechanisms. In \mbox{GdSb$_{0.46}$Te$_{1.48}$}, we believe that these two mechanisms are both active, which lays the foundation for a possible existence of a skyrmion phase. Below we list three aspects of magnetic properties that are similar to these three centrosymmetric skyrmion materials, especially \mbox{GdRu$_2$Si$_2$}.

Firstly, all these systems feature the large local magnetic moment from Gd with \textit{J} = 7/2, which is expected to be minimally affected by thermal fluctuations. This suggests that thermal fluctuations may not play as important of a role as in chiral magnets. Indeed the region of interest in the \textit{H-T} phase diagram for all these systems is located at relatively low temperatures (T < $\sim$20\,K for \mbox{Gd$_2$PdSi$_3$}, T < $\sim$13\,K for \mbox{Gd$_3$Ru$_4$Al$_{12}$}, T < $\sim$20\,K for \mbox{GdRu$_2$Si$_2$}, and T < 6.7\,K for \mbox{GdSb$_{0.46}$Te$_{1.48}$}). Theoretically, it was found that the skyrmion phase in a metallic centrosymmetric material can be stabilized as a ground state (\textit{T} = 0) by contributions that are higher-order than the conventional RKKY interactions \cite{ozawa2017zero}.

Secondly, all these systems are itinerant, and they show frustrations to a certain degree. In a square lattice, frustration comes from the \textit{nn}-interaction competing with additional longer-range-interaction (including the \textit{nnn}-interactions) \cite{lin2016ginzburg}. Fundamentally, the RKKY coupling, which oscillates as a function of Gd–Gd distance, is likely the origin of the frustration, similar to that in \mbox{Gd$_3$Ru$_4$Al$_{12}$} \cite{nakamura2018spin, hirschberger2019skyrmion}, \mbox{Gd$_2$PdSi$_3$} \cite{kurumaji2019skyrmion} and \mbox{GdRu$_2$Si$_2$} \cite{slaski1984magnetic, garnier1996giant}. The competition in magnetic interactions lead to the formation of many magnetic phases in the \textit{H-T} phase diagram, as was demonstrated in all four systems. A skyrmion phase in almost all known bulk materials features a narrow phase region, which is neighboring multiple magnetic phases in the \textit{H-T} diagram
. The small energy difference between these neighboring phases lays the foundation for the formation of skyrmion phase. 

Thirdly, superstructures are found in the zero-field magnetic ordered phases in all four systems. In the case of \mbox{Gd$_2$PdSi$_3$}, the crystal structure itself is characterized by a 2$\times$2 enlargement of the unit cell in the hexagonal basal plane and an eight-fold enlargement along the out-of-plane direction \cite{tang2011crystallographic, kurumaji2019skyrmion}. The ground-state magnetic order is characterized by an incommensurate noncoplanar spin texture. In the case of \mbox{Gd$_3$Ru$_4$Al$_{12}$}, spins form an incommensurate helical order at the ground state \cite{hirschberger2019skyrmion}. In the case of \mbox{GdRu$_2$Si$_2$}, an incommensurate screw magnetic structure was determined as the ground state \cite{khanh2020nanometric}. For \mbox{GdSb$_{0.46}$Te$_{1.48}$}, a single crystal XRD analysis indicated a five-fold superstructure along one of the in-plane lattice orientations. However, considering the possible limited \textit{q}-resolution of instrument and the tunable CDW wave vector in the \mbox{GdSb$_x$Te$_{2-x-\delta}$} series \cite{Lei2019Charge}, an incommensurate structure is also possible. With strong magnetoelastic coupling, the ground-state AFM1 phase is likely to have an incommensurate noncollinear magnetic order, as mentioned above. 

In addition to the aforementioned similarities to some known skyrmion materials, \mbox{GdSb$_{0.46}$Te$_{1.48}$} also exhibits some distinct magnetic properties. The neighboring phases of the anomalous region in \mbox{GdSb$_{0.46}$Te$_{1.48}$} [Fig. \ref{fig:GdSb0.46_low_field}(d)] are antiferromagnetic, in contrast to the typical field-aligned ferromagnetic (FA-FM) or paramagnetic (PM) neighboring phases in many skyrmion materials including centrosymmetric and non-centrosymmetric ones. Examples include the aforementioned \mbox{Gd$_2$PdSi$_3$} (PM) \cite{kurumaji2019skyrmion}, \mbox{FeGe} (FA-FM) \cite{yu2011near}, \mbox{MnSi} (FA-FM or PM) \cite{muhlbauer2009skyrmion}, and \mbox{Co$_8$Zn$_9$Mn$_3$} (PM) \cite{tokunaga2015new, bocarsly2019deciphering}, as well as other compounds, such as \mbox{VOSe$_2$O$_5$}  (PM) \cite{kurumaji2017neel}, \mbox{Fe$_{0.5}$Co$_{0.5}$Si} (FA-FM) \cite{yu2010real}, and \mbox{Cu$_2$OSeO$_3$} (PM) \cite{seki2012observation, levatic2014dissipation}) with well-defined phase diagrams. Therefore, if a skyrmion phase in \mbox{GdSb$_{0.46}$Te$_{1.48}$} is present, it would likely be of different type, for example it could be an antiferromagnetic skyrmion phase.

The concept of antiferromagnetic skyrmions was developed a few years ago \cite{zhang2016antiferromagnetic, barker2016static}, but it has been mainly limited to theoretical studies so far \cite{velkov2016phenomenology, jin2016dynamics, fujita2017ultrafast, gobel2017antiferromagnetic, salimath2020controlling}. In fact, antiferromagnetic skyrmions have not been experimentally observed in bulk materials yet, to the best of our knowledge. The difficulty in identifying promising materials that host antiferromagnetic skyrmions might be related to the vanishing skyrmion Hall effect, as is concluded from both the analytical theory and micromagnetic simulations \cite{zhang2016magnetic, velkov2016phenomenology, gobel2017antiferromagnetic, barker2016static, zhang2016antiferromagnetic, jin2016dynamics}. Such behavior prevents an efficient screening of candidate materials by electrical transport measurements. Antiferromagnetic skyrmions can be considered as two coupled ferromagnetic skyrmions with opposite topological winding numbers due to the presence of antiferromagnetic exchange interactions. Therefore, the winding number of an antiferromagnetic skyrmion is zero. Since the antiferromagnetic exchange interaction is a necessary ingredient, antiferromagnets are naturally the candidates in which to look for antiferromagnetic skyrmions. Based on the discussion above, we expect \mbox{GdSb$_{0.46}$Te$_{1.48}$} to be a promising candidate.



\subsection{\label{subsec2} Magnetic properties of \mbox{GdSb$_x$Te$_{2-x-\delta}$} with other compositions}

To understand the evolution of the magnetic phases with changing Sb content, we additionally studied \mbox{GdSb$_x$Te$_{2-x-\delta}$} compounds with Sb contents below and above 0.46.

\begin{figure*}
\includegraphics[width=1\textwidth]{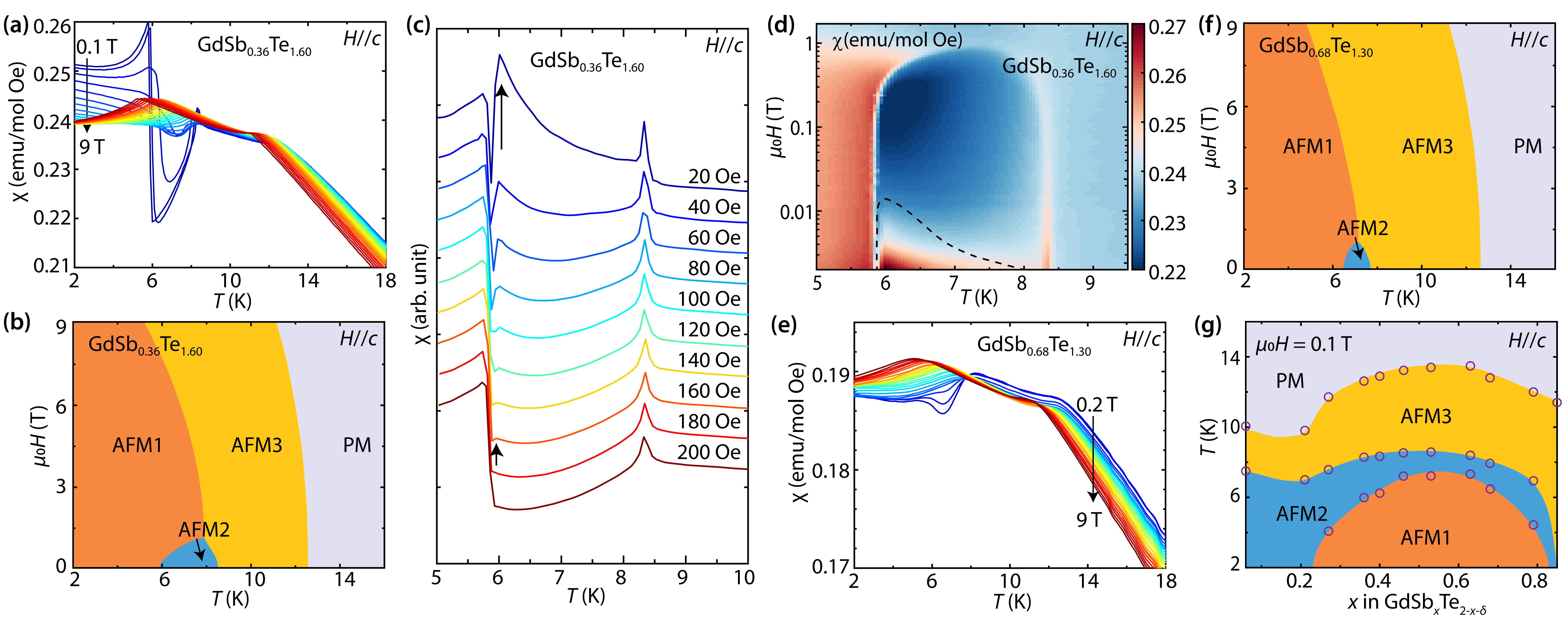}
\caption{\label{fig:GdSb0.36}Magnetic properties of \mbox{GdSb$_x$Te$_{2-x-\delta}$} with varying composition. (a) Temperature dependent DC magnetic susceptibility of \mbox{GdSb$_{1.36}$Te$_{1.60}$} at a selected field range of 0.1\,T - 9\,T, and (b) the corresponding phase diagram. (c) Temperature dependent DC magnetic susceptibility of \mbox{GdSb$_{1.36}$Te$_{1.60}$} at low fields from 20\,Oe to 200\,Oe, and (d) the corresponding magnetic susceptibility map. Arrows in (c) indicate anomalously enhanced $\chi_c$. The feature maintains for fields up to 180 Oe. The region with anomalously enhanced $\chi_c$ in (d) is outlined by the dashed line. (e) Temperature dependent DC magnetic susceptibility of \mbox{GdSb$_{1.68}$Te$_{1.30}$} at a selected field range of 0.1\,T - 9\,T, and (f) the corresponding phase diagram. (g) Evolution of the low-field magnetic phases with respect to Sb composition. AFM1, AFM2, and AFM3 indicate three antiferromagnetic phases, as discussed in the main text. The  circles indicate the three phase transition boundaries extracted from $\chi_c\left(T\right)$ at $\mu_0H\,=\,0.1\,T$. Note that the possible skyrmion phase observed in \mbox{GdSb$_{0.46}$Te$_{1.48}$} is not marked in the diagram, as the anomalous response in $\chi_c$ was not observed in the compounds with neighboring Sb compositions (\mbox{GdSb$_{0.40}$Te$_{1.54}$}, and \mbox{GdSb$_{0.53}$Te$_{1.44}$}) when $\mu_0H$\,=\,0.1\,T.}
\end{figure*}

Figures \ref{fig:GdSb0.36}(a) and (b) show the temperature dependent DC magnetic susceptibilities under fields up to 9\,T (\textit{H//c}) and the magnetic phase diagram of \mbox{GdSb$_{0.36}$Te$_{1.60}$}, respectively. Overall, \mbox{GdSb$_{0.36}$Te$_{1.60}$} has similar magnetic properties to \mbox{GdSb$_{0.46}$Te$_{1.48}$}. Three magnetic phases: AFM1, AFM2, and AFM3, can be identified. The shapes of these phase regions in the \textit{H-T} diagram is similar to those of \mbox{GdSb$_{0.46}$Te$_{1.48}$} [Fig. \ref{fig:GdSb0.46_Hc}(d)], except that the phase region of AFM2 moves to slightly lower temperature, and it exists at a wider temperature range at low fields. 
For the AFM1 phase, we found that the magnetic susceptibility at low-field ($\mu_0H=0.1$\,T) is significantly larger than that at high field, which is different to \mbox{GdSb$_{0.46}$Te$_{1.48}$}. This indicates a stronger polarizability along the \textit{c}-axis for the low-field AFM1 phase of \mbox{GdSb$_{0.36}$Te$_{1.60}$} compared to the same phase in \mbox{GdSb$_{0.46}$Te$_{1.48}$}.

Notably is the low-field magnetic behavior of \mbox{GdSb$_{0.36}$Te$_{1.60}$} in comparison with \mbox{GdSb$_{0.46}$Te$_{1.48}$}. Figure \ref{fig:GdSb0.36}(c) shows the temperature dependent DC magnetic susceptibility under fields up to 200\,Oe. Within the AFM2 phase region, anomalous peaks can be resolved for fields up to 180\,Oe. This region is highlighted in Fig. \ref{fig:GdSb0.36}(d). The existence of such a region with enhanced low-field susceptibility might share the same origin as that in \mbox{GdSb$_{0.36}$Te$_{1.60}$}, although the former region is within the AFM2 phase while the latter one is within the AFM1 phase. The appearance of such a region in two different phases indicates a change of energy landscape in magnetic interactions, upon a change in Sb/Te ratio. 

Moving to compounds with higher Sb composition, we show the temperature dependent DC magnetic susceptibilities and the \textit{H-T} phase diagram of \mbox{GdSb$_{0.68}$Te$_{1.30}$} in Figs. \ref{fig:GdSb0.36}(e) and (f), respectively. Although all three magnetic phases are still observed, the phase space of AFM2 is getting smaller. Additionally, the low-field magnetic susceptibility of the AFM1 phase is clearly smaller than the high-field one [Fig. \ref{fig:GdSb0.36}(e)]. We do not observe any regions with anomalously enhanced low-field susceptibility in \mbox{GdSb$_{0.68}$Te$_{1.30}$}, in contrast to that in \mbox{GdSb$_{0.36}$Te$_{1.60}$} and \mbox{GdSb$_{0.46}$Te$_{1.48}$}. 

In Fig. \ref{fig:GdSb0.36}(g), we summarize the magnetic phase evolution in respect to the Sb content (\textit{x}) ($\mu_0H\,=0.1\,T$). The phase regions of AFM1, AFM2, and AFM3 are all dome-shaped, and the phase transition temperatures summit when $x\sim0.55-0.60$. Additionally, the AFM1 phase disappears at both the low-\textit{x} and high-\textit{x} limit. 

The increase of the AFM3-AFM2 and AFM2-AFM1 transition temperatures upon decrease in \textit{x} on the high-\textit{x} side of the $T-x$ diagram is related to the development of CDW distortion that breaks the $C_4$ symmetry, and the gradual increase of $a/b$ ratio 
\cite{Lei2019Charge}. In this sense, the formation of the AFM1 and AFM2 phases should be related to the emergence of an in-plane magnetocrystalline anisotropy when $x<0.85$. However, the orthorhombic distortion might not be the only factor that affects the stability of the AFM1 and AFM2 phases, because the maximum transition temperatures appear at $x\sim0.55-0.60$, while the maximum \textit{a/b}-ratio occurs at $x\sim0.25$ \cite{Lei2019Charge}. This implies the contributions from other energy terms. For a metallic antiferromagnet, this could be from long-range RKKY interactions. The RKKY interaction is sensitive to the density of states because it depends on the number of electrons contributing to the spin exchange. With decrease in \textit{x}, the crystal structure of \mbox{GdSb$_x$Te$_{2-x-\delta}$} tends to distort more, eventually leading to a reduction in the number of conducting electrons to interact with the local spins, which likely disfavors the formation of AFM1 and AFM2 phases.  

On the low-\textit{x} ($x<\sim$0.2) side of the phase diagram shown in Fig. \ref{fig:GdSb0.36}(g), the PM-AFM3 and AFM3-AFM2 transition temperatures appear to slightly increase with lower \textit{x}. This might be related to the structural transition from a unidirectional CDW distortion to a bi-directional one, when \textit{x} is reduced to below $\sim0.2$ \cite{Lei2019Charge}.

\subsection{\label{subsec3} Magnetic properties of \mbox{GdTe$_3$}}

As discussed above, the in-plane magnetic exchange interactions, including two anisotropic \textit{nn}-interactions (\textit{J$_{1a}$} and \textit{J$_{1b}$}) and the \textit{nnn}-nearest interaction (\textit{J$_{2}$}) shown in Fig. \ref{fig:structure}, are responsible for the many magnetic ordered phases observed in \mbox{GdSb$_x$Te$_{2-x-\delta}$}. However, the out-of-plane magnetic exchange interactions (\textit{J$_{c}$} and \textit{$J^\prime_{c}$}), could also play a role. Therefore, we resort to a structurally related material, \mbox{GdTe$_3$}.

\begin{figure}
\includegraphics[width=0.5\textwidth]{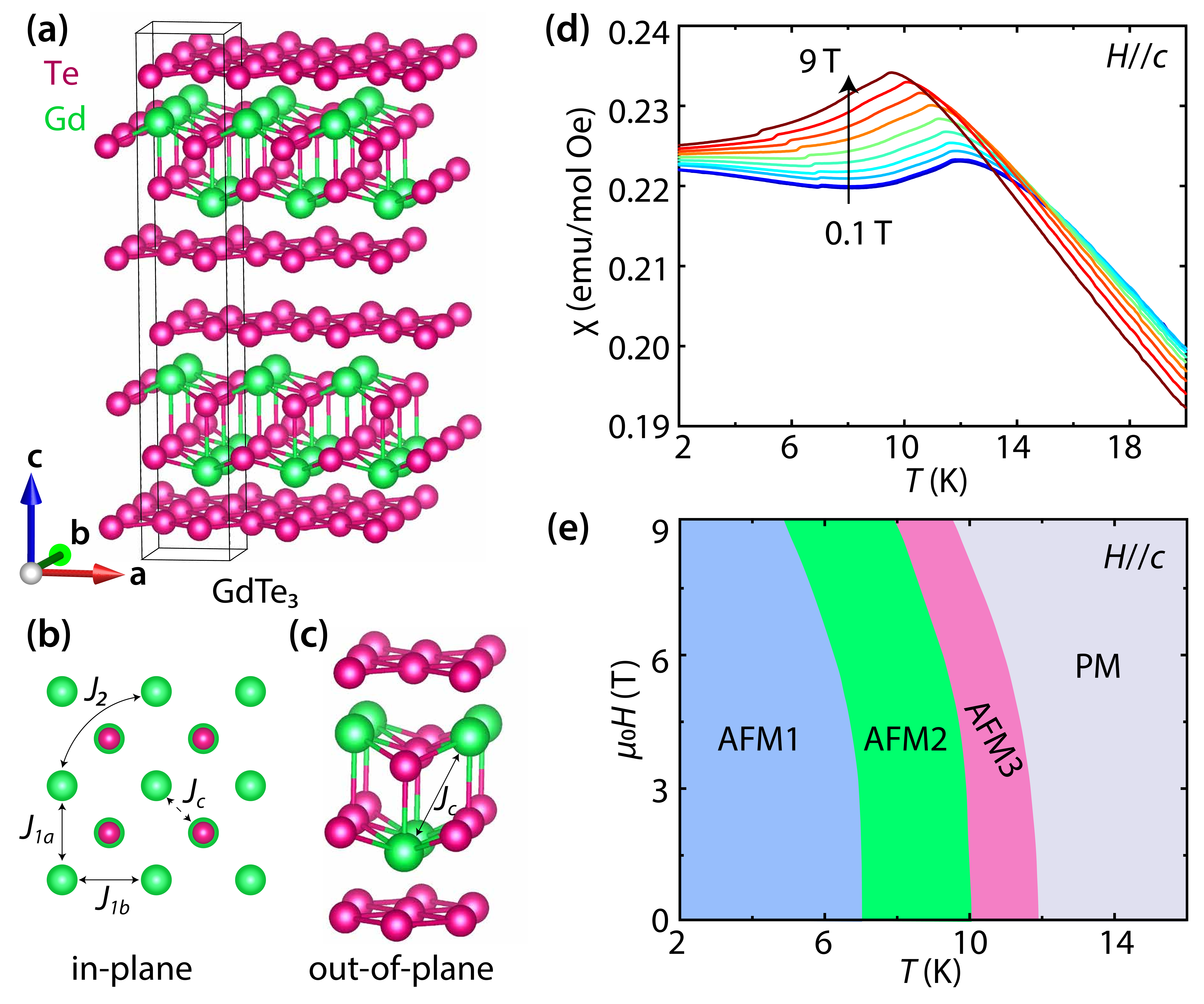}
\caption{\label{Figure7}Magnetic phase diagram of \mbox{GdTe$_3$}. (a) Illustration of crystal structure of \mbox{GdTe$_3$}. In-plane (b) and out-of-plane (c) exchange coupling terms. (d) Temperature dependent DC magnetic susceptibility at a selected field range of 0.1 T - 9 T, for fields along the out-of-plane orientation. 
(e) Magnetic phase diagrams for $H//c$. PM indicates paramagnetic phase. AFM1, AFM2, and AFM3 are antiferromagnetic phases, as discussed in the main text.}
\end{figure} 

The structure of \mbox{GdTe$_3$}, without accounting for the CDW, is illustrated in Fig. \ref{Figure7}(a). Just as \mbox{GdSb$_x$Te$_{2-x-\delta}$} ($\sim0.20\,<x\,<0.85$), \mbox{GdTe$_3$} also features unidirectional-CDW-distorted square lattices. Therefore, we expect similar in-plane magnetic exchange interaction terms [Fig. \ref{Figure7}(b)] as in \mbox{GdSb$_x$Te$_{2-x-\delta}$}. However, the structural difference along the out-of-plane direction should lead to different out-of-plane exchange interactions. Each GdTe bilayer is separated by two square-lattice planes in \mbox{GdTe$_3$}, but only one in \mbox{GdSb$_x$Te$_{2-x-\delta}$}. Therefore, the interbilayer magnetic exchange interaction ($J^\prime_{c}$) in \mbox{GdTe$_3$} is expected to be significantly weaker than that in \mbox{GdSb$_x$Te$_{2-x-\delta}$}, and only the intrabilayer out-of-plane exchange interaction (\textit{J$_{c}$}) is considered to be a dominant term [Fig. \ref{Figure7}(c)].

Figures \ref{Figure7}(d) and (e) show the temperature dependent DC magnetic susceptibility [$\chi_c\left(T\right)$] and the \textit{H-T} phase diagram for $H//c$. The low-field susceptibility $\chi_c\left(T\right)$ remains fairly flat down to the lowest measured temperature below $T_N$, which is different from \mbox{GdSb$_{0.46}$Te$_{1.48}$}, where the susceptibility dips [Fig.  \ref{fig:GdSb0.46_Hc}(a)]. Therefore, at least the AFM2 phase should be different between these two materials. We expect an in-plane magnetic easy axis for all three magnetically ordered phases in \mbox{GdTe$_3$}.


The difference in the magnetic phase diagrams between \mbox{GdTe$_3$} and \mbox{GdSb$_x$Te$_{2-x-\delta}$} is linked to the role of the interbilayer out-of-plane exchange coupling term, \textit{$J^\prime_{c}$}, in \mbox{GdSb$_x$Te$_{2-x-\delta}$}. In the discussion above, we have considered the roles of the in-plane \textit{nnn}-interactions in competition with the \textit{nn}-interaction term in the formation of the complex magnetic phases. Based on the comparison with GdTe$_3$, it can be concluded that the out-of-plane exchange interaction, particularly the interbilayer out-of-plane exchange coupling term also plays an important role for the formation of magnetic phases with an out-of-plane easy axis, and therefore also affects the potential existence of a skyrmion phase.

\section{Conclusion and Outlook}


In this work, we have reported the magnetic properties of square-lattice based, topological semimetal candidates \mbox{GdSb$_x$Te$_{2-x-\delta}$}. In the orthorhombic phase, we reported rich magnetic phase diagrams for three different field orientations. At low field, we found an anomalous region in the \textit{H-T} phase diagram with enhanced magnetic susceptibility, which suggests the existence of antiferromagnetic skyrmions. To understand the rich magnetic phases, we analyzed the roles of \textit{nn}- and \textit{nnn}- in-plane magnetic exchange interactions, and intrabilayer and interbialyer out-of-plane exchange interactions. An active interplay of all these interactions was found to be responsible for the observed plethora in magnetic phases. Frustrated RKKY interactions and higher-order coupling beyond the RKKY interactions are the two fundamental motifs contributing to the competition of these terms in \mbox{GdSb$_x$Te$_{2-x-\delta}$} and might lead to an antiferromagnetic skyrmion phase in \mbox{GdSb$_{0.46}$Te$_{1.48}$}. In this context, we would like to mention \mbox{\textit{R}Mn$_2$Ge$_2$} ($R\,=\,$La, Ce, Pr, and Nd) with a Mn square lattice. Here, the Mn$^{2+}$ forms a high spin state that is largely localized. In this system, it was argued that the long-range RKKY interactions via mobile electrons are a primary source for the out-of-plane Mn-Mn coupling, giving rise to frustrated interactions and the conical spin order \cite{kolmakova2002features}.

More broadly, we believe the square-lattice system of \mbox{\textit{Ln}Sb$_x$Te$_{2-x-\delta}$} presents an outstanding platform to investigate the rich physics endowed by complex magnetism, CDWs, Dirac semimetal states, and their possible interplay. The possible existence of an antiferromagnetic skyrmion in the same phase that was shown to be an ideal non-symmorphic Dirac semimetal, makes this family of materials even more appealing, as antiferromagnetic skyrmions have not been experimentally observed in any bulk materials before. Our results therefore point to the existence of antiferromagnetic skyrmions in an ideal Dirac semimetal. 

\begin{acknowledgments}
This research was supported by the Arnold and Mabel Beckman Foundation through a Beckman Young Investigator grant awarded to L.M.S, and the Princeton Center for Complex Materials, a National Science Foundation (NSF)-MRSEC program (DMR-1420541). The authors acknowledge the use of Princeton’s Imaging and Analysis Center, which is partially supported by the Princeton Center for Complex Materials.
\end{acknowledgments}



\providecommand{\noopsort}[1]{}\providecommand{\singleletter}[1]{#1}%

\end{document}



\title{Supporting Information for ``Complex magnetic phases enriched by charge density waves in topological semimetals GdSb$_x$Te$_{2-x-\delta}$''}

\author{Shiming Lei}
\affiliation{Department of Chemistry, Princeton University, Princeton, New Jersey 08544, USA}

\author{Audrey Saltzman}
\affiliation{Department of Physics, Massachusetts Institute of Technology, Cambridge, MA 02139, USA}


\author{Leslie M. Schoop}
 \email{lschoop@princeton.edu}
\affiliation{Department of Chemistry, Princeton University, Princeton, New Jersey 08544, USA}

\date{\today}

\maketitle



\subsection{\label{section1}MH-data decoupling}
\noindent As mentioned in the main text, the measured magnetization can be considered as a simple summation of the magnetization for two species of domains in which the field is either applied parallel to \textit{a} or \textit{b}. If we define $\lambda$ as the volume fraction of the \textit{a}-domain, in which the \textit{a}-axis is parallel to field orientation, then the intrinsic magnetic responses $M_a$(\textit{H}) and $M_b$(\textit{H}) can be described by eqs. (3) and (4).
\\
\\
\noindent Once $\lambda$ is known, the intrinsic magnetic response without domain mixing can be obtained also if the measurements were performed on a crystal with ferroelastic twin domains. To determine $\lambda$, the features visible in the field-dependent magnetic susceptibility can be utilized. For an antiferromagnetic phase, a spin-flop transition can appear at one specific critical field under a specific field orientation. We can use this feature to deconvolute the domain mixing effect. For the field induced transitions ($\mu_0H\approx  1.7\,$T and 5.7\,T) shown in Fig. 4(a) and (b), the transition at 1.7\,T is more pronounced for \textit{H//IP1} than \textit{H//IP2}, while the transition at 5.7\,T is more pronounced for \textit{H//IP2} than \textit{H//IP1}. We conclude that the weaker transition features, which appear at 5.7\,T for \textit{H//IP1} and at 1.7\,T for \textit{H//IP2} [indicated by the red arrows in Fig. 4(a) and (b)], come from the domain mixing.
\\
\\
Numerically we continuously tune the values of $\lambda$ in eqs. (3) and (4) from 0 to 1 and monitor the evolution of the resulting $M_a\left(H\right)$ and $M_b\left(H\right)$. Upon a correct input of $\lambda$, the weaker features shown in Fig. 4(a) and (b) are expected to diminish, while the dominant transition features [indicated by the black arrows in Fig. 4(a) and (b)] are retained. Our numerical tests indicate that when $\lambda$ =  0.75, the transition at 1.7\,T is only visible in $M_a\left(H\right)$ [Fig. 4(c)], but not in $M_b\left(H\right)$ [Fig. 4(d)]. In the meanwhile, the feature at 5.7\,T is almost only seen in $M_b\left(H\right)$. Therefore, the field dependent magnetic susceptibility shown in Fig. 4(c) and (d) is considered as the intrinsic magnetic response without domain mixing.

\subsection{\label{section2}Mechanical detwinning of \mbox{GdSb$_{0.53}$Te$_{1.44}$} crystals}
\noindent For mechanical detwinning, a \mbox{GdSb$_{0.53}$Te$_{1.44}$} crystal was clamped by two metallic plates inside a vacuum-sealed quartz tube. Figure \ref{figureS2}(a) shows an illustration of mechanical detwinning by an external compressive stress. The sealed quartz tube was slowly heated to $600\,^\circ$C to facilitate the detwinning, followed by cooling over 6 hours to room temperature. We found that a slow heating and cooling is important to minimize cracks in the detwinned crystals. Figure \ref{figureS2}(b) shows polarized optical images before and after detwinning. The disappearance of the bright-dark contrast of the stripe-like domain suggests successful detwinning. Since the lattice parameter \textit{a} along CDW wave vector is larger than \textit{b}, the $a$-axis is aligned along the horizontal direction after detwinning. To evaluate the success of detwinning, further isothermal magnetization [Fig. \ref{figureS2}(c)] and temperature dependent magnetization [Fig. \ref{figureS2}(d)] were measured for \textit{H//a}. The spin-flop transition at $\mu_0H=5.6$\,T is clearly suppressed in the field dependent magnetization measured after detwinning. Therefore, we conclude that the spin-flop transition at $\mu_0H=1.8$\,T is intrinsic to field condition of \textit{H//a}, while the transition at $\mu_0H=5.6$T is related to \textit{H//b}. Based on this result, we are able to specify the field orientation for the magnetic phase diagrams shown in Fig. 4 in the main text. We noticed a slight residue of curvature at $\mu_0H=5.6$\,T in the field dependent magnetization measured after detwinning, which implies the persistence of a small fraction of twin domains. The residual twin domains might be buried inside the crystals, and can not be observed by polarized optical imaging.

\clearpage

\begin{figure*}[h]
  \includegraphics[width=0.8\textwidth]{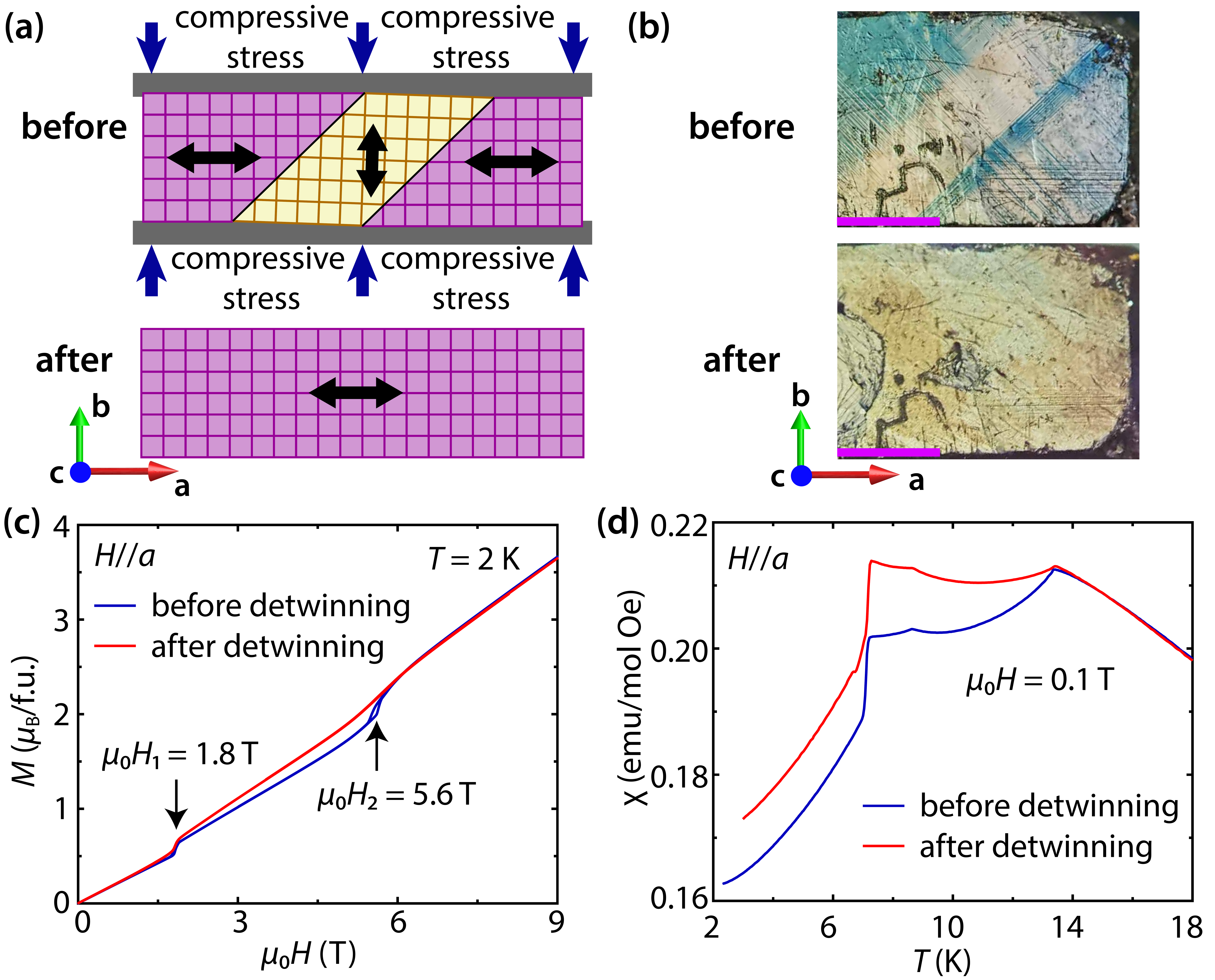}
\caption{Mechanical detwinning of \mbox{GdSb$_{0.53}$Te$_{1.44}$} crystals.  (a) Illustration of mechanical detwinning of crystals by an external uniaxial compressive stress. The double-headed arrows indicate the CDW wave vector orientation. Different shades of color indicate different ferroelastic domains. (b) Polarized optical image of a \mbox{GdSb$_{0.53}$Te$_{1.44}$} crystal before and after detwinning. Scale bar at the bottom left corner of each image is $500\,\mu$m. (c) and (d) \textit{MH}-curves and temperature dependent magnetization measured on the GdSb$_{0.53}$Te$_{1.44}$ crystal before and after detwinning. Arrows in (c) indicate the locations of two spin-flop transitions.}
\label{figureS1}
\end{figure*}

\clearpage
\subsection{Temperature dependent in-plane magnetic susceptibility for \mbox{GdSb$_{0.46}$Te$_{1.48}$}}
\begin{figure*}[h]
  \includegraphics[width=0.6\textwidth]{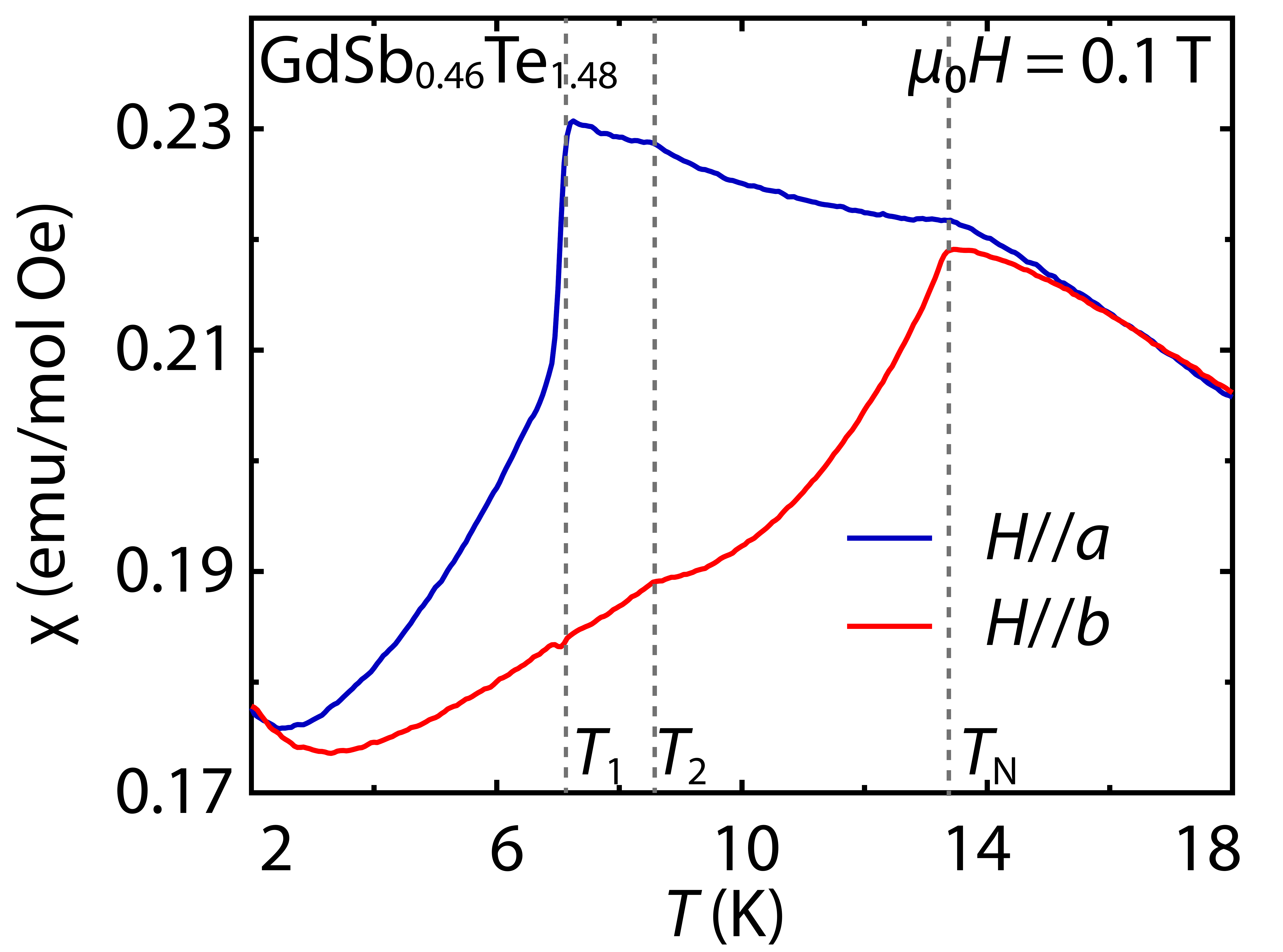}
\caption{Decoupled temperature dependent magnetic susceptibility for \mbox{GdSb$_{0.46}$Te$_{1.48}$}, for \textit{H//a}, and \textit{H//b}. Dashed lines indicate the three magnetic transition temperatures. Measurements were performed under a DC field of 0.1\,T. The decoupling procedure is described in the main text.}
\label{figureS2}
\end{figure*}

\clearpage
\subsection{\label{section1}H-T phase diagram of \mbox{GdSb$_{0.63}$Te$_{1.36}$} for \textit{H//a} and \textit{H//b}}
\begin{figure*}[!h]
\includegraphics[width=0.65\textwidth]{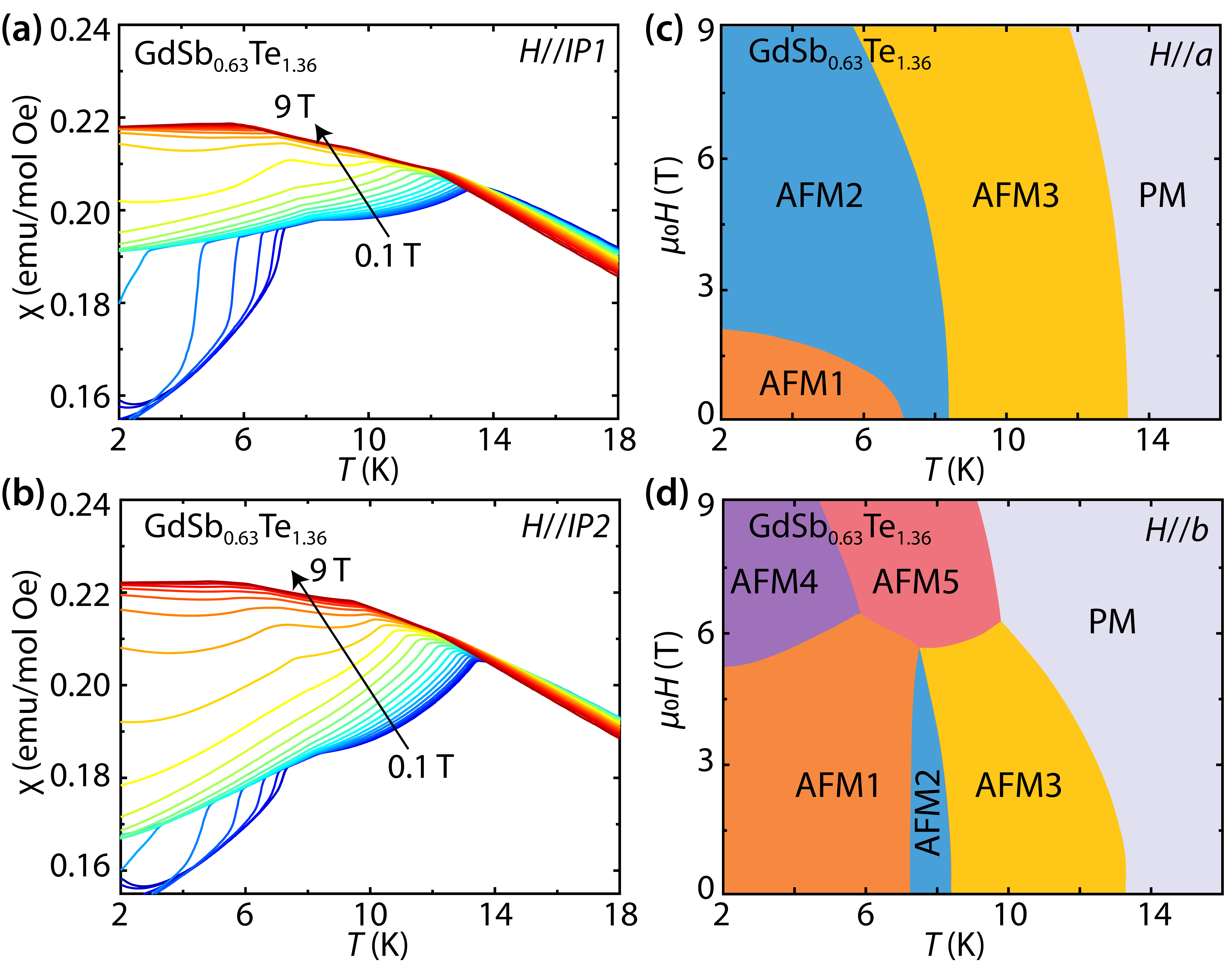}
\caption{\label{FigureS3}Magnetic phase diagram of \mbox{GdSb$_{0.63}$Te$_{1.36}$} for field parallel to the in-plane orientations. (a) and (b) Temperature dependent DC magnetic susceptibility at selected field range from 0.1\,T to 9\,T, for \textit{H//IP1} and \textit{H//IP2}, respectively. (c) and (d) Magnetic phase diagrams for \textit{H//a} and \textit{H//b}, respectively. PM indicates the paramagnetic phase. AFM1, AFM2, and AFM3 are antiferromagnetic phases, as discussed in the main text.}
\end{figure*}

\noindent Figures \ref{FigureS3}(a) and (b) show the magnetic susceptibility maps for the two in-plane orientations, \textit{IP1} and \textit{IP2}, respectively. The magnetic anisotropy is evident, similar to that of \mbox{GdSb$_{0.46}$Te$_{1.48}$} [Figs. 4(a) and (b)]. Figures \ref{FigureS3}(c) and (d) illustrate the \textit{H-T} phase diagrams for \textit{H//a} and \textit{H//b}, respectively. The magnetic phase diagrams are similar to that of \mbox{GdSb$_{0.46}$Te$_{1.48}$} [Fig. 4(g) and (h)]. One noticeable change in \mbox{GdSb$_{0.63}$Te$_{1.36}$}, however, is the enlargement of the phase region of AFM4, compared to \mbox{GdSb$_{0.46}$Te$_{1.48}$}: the AFM4-AFM1 phase boundary is pushed to a lower field, and the AFM4-AFM5 phase boundary is pushed to a higher temperature. With further increase of \textit{x} in \mbox{GdSb$_x$Te$_{2-x-\delta}$}, the AFM4-AFM1 phase boundary is expected to move towards lower fields until the disappearance of AFM1 phase, in accordance with \textit{T-x} phase diagram shown in Fig. 6(g).
\\
\\

\bibliography{Reference}